# Coherent population trapping and spin relaxation of a silicon vacancy center in diamond at mK temperatures

Shuhao Wu[1], Xinzhu Li[1], Ian Gallagher[2], Benjamin Lawrie[2,#], and Hailin Wang[1,*]

[1]Department of Physics, University of Oregon, Eugene, OR 97403, USA

[2]Materials Science and Technology Division, Oak Ridge National Laboratory, Oak Ridge, TN 37831

Abstract

We report experimental studies of coherent population trapping and spin relaxation in a temperature range between 4 K and 100 mK in a silicon vacancy (SiV) center subject to a transverse magnetic field. Near and below 1 K, phonon-induced spin dephasing becomes negligible compared with that induced by the spin bath of naturally abundant $^{13}$C atoms. The temperature dependence of the spin dephasing rates agrees with the theoretical expectation that phonon-induced spin dephasing arises primarily from orbital relaxation induced by first order electron-phonon interactions. A nearly 100-fold increase in spin lifetime is observed when the temperature is lowered from 4 K to slightly below 1 K, indicating that two-phonon spin-flip transitions play an essential role in the spin relaxation of SiV ground states.

#lawriebj@ornl.gov
*hailin@uoregon.edu

I. Introduction

Color centers in diamond have emerged as a promising qubit platform for quantum information processing[1-4]. While negatively charged nitrogen vacancy (NV) centers in diamond feature long spin decoherence times for electron spins even at room temperature [1-3], negatively charged SiV centers feature superior optical properties [4-7]. Because of their inversion symmetry, SiV centers feature optical transitions that are robust against charge fluctuations[5-7]. The zero-phonon line of SiV centers contains about 70% of the total fluorescence, with a nearly lifetime-limited optical linewidth. Excellent optical coherence has been observed for SiV centers in diamond nanostructures and in diamond membranes as thin as 100 nm[8-12]. All-optical as well as microwave control of SiV spins has been demonstrated [13-16]. In addition, the integral orbital degrees of freedom in the SiV ground spin states make it feasible to control the SiV spin with mechanical vibrations[17, 18], which also suggests the possibility of quantum control of an electron spin at the level of single phonons in a diamond spin-mechanical resonator. Strong orbital interactions with thermal phonons, however, lead to relatively fast spin dephasing and spin relaxation at elevated temperatures.

Spin dephasing, specifically $T_2^*$, of SiV centers has been investigated with Ramsey interferometry as well as coherent population trapping (CPT) [13-16], though there have been no reports of CPT studies below 4 K. Phonon-induced spin dephasing can be suppressed with decreasing temperatures. Nearly complete suppression of phonon-induced dephasing has been observed at temperatures near 100 mK, where decoherence times can exceed 10 ms in isotopically enriched diamond [19]. A relatively large ground-state orbital splitting, which can be achieved in group IV color centers such as tin vacancy or induced by an externally applied mechanical strain, can also effectively suppress the effects of thermal phonons[20, 21]. In addition to spin dephasing studies, spin relaxation has also been investigated with spin lifetime measurements at temperatures near 4 K and at 100 mK[13-16]. Additional experimental studies between 4 K and 100 mK can provide further information on phonon-induced spin dephasing and relaxation processes.

The latest experimental advance on the realization of GHz diamond spin-mechanical resonators with mechanical Q-factors exceeding $10^6$ have stimulated considerable efforts in the experimental pursuit of phononic cavity QED with SiV spins[22-24]. For these studies, phonon-induced relaxation and CPT related processes at mK temperature are expected to play an important role. Recently, CPT has also been exploited for real-time quantum sensing with a NV center[25, 26]. CPT studies at mK temperatures will be an essential step in the use of CPT-based real-time sensing to suppress spin dephasing in SiV centers induced by the nuclear spin bath.

In this paper, we report experimental studies of CPT at temperatures as low as 100 mK and spin relaxation between 4 K and 830 mK in a SiV center subject to a transverse magnetic field. The CPT studies show that near and below 1 K, spin dephasing is primarily due to the spin bath of naturally abundant $^{13}$C atoms, with negligible contributions from phonon-induced spin dephasing. The temperature dependence of the spin dephasing rates agrees with the theoretical expectation that the phonon-induced spin dephasing arises primarily from orbital relaxation of the SiV ground states induced by the first order electron-phonon interactions. In addition, a nearly 100-fold increase in spin lifetime is observed when the temperature is lowered from 4 K to slightly below 1K. A numerical analysis of the observed temperature dependence of the spin lifetime along with a qualitative consideration of relevant spin relaxation processes in the SiV ground states reveals that two-phonon spin-flip transitions play an essential role in the spin relaxation of the SiV ground states.

II. Sample and experimental setup

Our experimental studies were carried out with a SiV center implanted about 75 nm below the surface of an electronic grade diamond grown by chemical vapor deposition (CVD). The average kinetic energy and dosage of the $^{28}$Si used in the implantation are 100keV and $3 \times 10^9$/cm$^2$, respectively. Stepwise thermal annealing up to a temperature of 1200 degrees followed by wet chemical oxidation was used for the removal of the damaged surface layer as well as for the formation of SiV centers. The sample was thermally anchored to the mixing chamber of a dilution refrigerator (Leiden Cryogenics CFCS81-1000M), which is fitted with a 3D vector magnet (American Magnetics). The SiV fluorescence was collected with an objective with NA=0.82, followed by an optical collimation system installed in a cold-insertable probe in the refrigerator[27]. SiV fluorescence exiting an optical window at the top of the refrigerator was coupled into a multi-mode fiber with a diameter of 10 μm and then sent to an avalanche photodiode for photon counting. For photoluminescence excitation (PLE) spectra, a 532 nm laser pulse was used for the initialization of the SiV center. A red laser with a wavelength near 737 nm (New Focus Velocity TLB-6700) was used for the resonant excitation of the SiV center. An electro-optic modulator (EOM) was used to generate the two optical fields needed for the CPT experiments.

One field came from the carrier wave and the other came from the first sideband generated by the EOM. Optical pulses needed were generated with acousto-optic modulators (AOMs). Time resolved florescence measurements were performed with a time tagger (Swabian Instruments).

III. Results and discussions

For a negatively charged SiV center, both the ground and excited states are characterized by orbital states, |$e_x$> and |$e_y$>, and spin states, |↑> and |↓>. Spin-orbit coupling leads to the formation of doubly degenerate doublets for the ground and excited states, with the two lower states being $|e_+,\uparrow>$ and $|e_-,\downarrow>$ and the two upper states being $|e_-,\uparrow>$ and $|e_+,\downarrow>$, where $|e_\pm> = (|e_x> \pm i|e_y>)/\sqrt{2}$, as shown schematically in Fig. 1a [28]. The splitting between the upper and lower states is $\lambda_{so}$, with ground-state splitting $\lambda_{so}^g$ near 50 GHz and excited-state splitting $\lambda_{so}^e$ near 260 GHz. In the absence of an external magnetic field, optical transitions between the ground and excited states are all spin-conserving. An off-axis magnetic field removes the spin degeneracy and induces spin-state mixing of the ground and excited states, as illustrated in Fig. 1a. The spin-state mixing enables spin-flip optical as well as microwave and acoustic transitions. Note that both the spin and orbital states are slightly mixed. The orbital state mixing arises from strain and Jahn-Teller effects [28]. As illustrated in Fig. 1b, a magnetic field with a direction normal to the SiV axis and with $B$=0.12 T was used for all the experimental results presented in this paper, unless otherwise specified, to maximize the spin-state mixing at the given field strength. Our experimental studies were carried out on the C-transition of the SiV center (see Fig. 1a). Figure 1c shows the PLE spectrum obtained for the C-transition at 4 K. The two resonances in the PLE spectrum correspond to the spin-conserved $C_2$ and $C_3$ transitions.

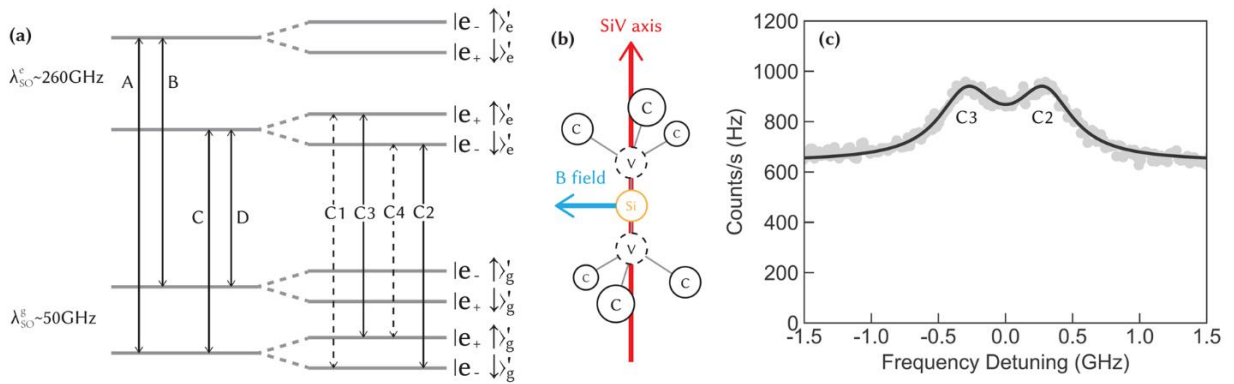

**Fig. 1.** (a) Energy level structure of the SiV center in the presence of an off-axis magnetic field. The prime in the state label indicates that the spin states as well as the orbital states are slightly mixed. (b) Schematic of a SiV center subject to a magnetic field normal to the SiV axis. (c) PLE spectrum of the SiV C-transition obtained at 4 K.

The spin-conserved $C_2$ transition and the spin-flip $C_4$ transition in Fig. 1a form a Λ-type three-level system for our CPT experiments. The spectral position of the weak spin-flip transition, however, is difficult to identify directly in the PLE spectrum. We searched for the weak spin-flip transition with two approaches, optical pumping and CPT, but with essentially the same experimental setting, for which an optical carrier wave is fixed at the spin conserved $C_2$ transition, while the frequency of the first sideband generated by the EOM is scanned over the expected spectral range of the spin-flip transition. After this set of experiments, we changed the experimental setting such that the optical carrier wave is fixed at the weak spin-flip $C_4$ transition and the first sideband is now near the spin-conserved transition. This is because strong CPT dips occur when the Rabi frequencies for the spin-conserved and spin-flip optical transitions are comparable. Figure 2a shows, as an example, a CPT spectral response obtained in this setting at 4 K. CPT spectral responses depend strongly on the incident optical powers, as shown in Fig. 2b. Figure 2c plots the CPT linewidth as a function of the incident optical power obtained at 150 mK. We extract the intrinsic CPT linewidth by fitting the power-broadened CPT linewidth to a linear power dependence. Note that the error bars plotted in Figs. 2c and 2d include uncertainties in the numerical fits and an additional 10% uncertainty due to polarization fluctuations in the experiment.

In a traditional or textbook CPT setting[29], a pump and a probe field are employed. The Rabi frequency for the probe is assumed to be sufficiently weak such that the optical response is linear with respect to the probe field. In this limit, the CPT linewidth scales linearly with the intensity of the pump field. We present in the appendix a detailed theoretical calculation on the power broadening of the CPT linewidth, when the optical response is nonlinear to both incident fields. To solve the problem analytically, we have assumed that the decay rate of the spin coherence is small compared with all other relevant decay rates of the system such that the excited-state population and the optical dipole coherences involved reach steady state in a timescale much shorter than the decay time of the spin coherence. With this assumption, we show that the CPT linewidth is proportional to the intensity of both incident laser fields. It should be noted that this result, while expected, is different from the well-known behavior of power broadening for a dipole optical transition.

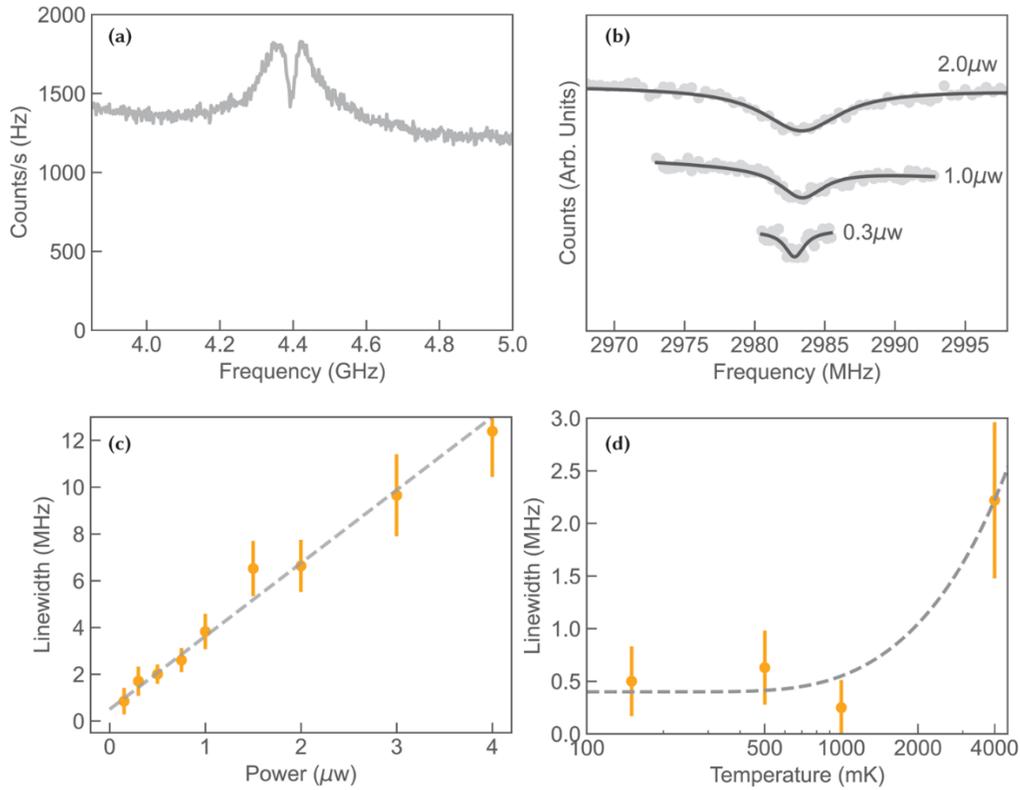

**Fig. 2.** (a) CPT spectral response obtained at 4K with an input laser power of 3 µW. (b) Power dependence of the CPT spectral response at 0.15 K. The incident optical powers used are indicated in the figure. (c) Power dependence of the CPT linewidth at 0.15 K. The linear fit yields an intrinsic CPT linewidth of $0.5 \pm 0.33$ MHz. (d) The intrinsic CPT linewidth as a function of temperature. The dashed line is a numerical fit, assuming that the phonon-induced dephasing rate is proportional to the thermal occupation of phonons with a frequency of 50 GHz. Except for the CPT spectral response in (a), which was obtained with $B= 0.15$ T along the crystal z-axis, all other data were obtained with $B= 0.12$ T normal to the SiV axis.

Figure 2d plots the intrinsic CPT linewidths derived from the linear power dependence as a function of temperature from 4 K to as low as 150 mK. Note that stable temperatures between 4 K and 1 K were difficult to achieve in our current setup. At 4 K, the spin dephasing time, $T_2^*$, derived from the intrinsic CPT linewidth (2.35 MHz) is in general agreement with that obtained from earlier Ramsey interferometry studies [16]. CPT linewidths at 4 K reported from earlier studies significantly exceed the intrinsic linewidth obtained in our experiment, which is likely due to power broadening of the CPT spectral response in the earlier experiments. At temperatures near and below 1 K, the intrinsic CPT linewidths are nearly independent of temperature, indicating that phonon-induced dephasing is negligible compared with dephasing induced by the nuclear spin bath.

Earlier experimental and theoretical studies have attributed the phonon-induced spin dephasing at temperatures near and above 4 K to the orbital relaxation between the two lower and upper states in the ground state doublet, induced by the first order electron-phonon interaction with phonon frequencies near $\lambda_{so}^g$ [30]. In this case, the phonon-induced spin dephasing rate is expected to be proportional to the thermal phonon occupation, $n(T) = 1/[\exp(h\lambda_{so}^g/k_BT) - 1]$. The numerical fit shown in Fig. 2d indicates a good agreement between the experimental result and the theoretical expectation for temperatures below 4 K.

At T< 1 K, thermal populations of phonons near 50 GHz are greatly suppressed. Phonon-induced spin dephasing and spin relaxation, however, can still arise from the direct acoustic transition between the two lower energy ground spins states, especially when a transverse magnetic field is used to maximize spin-state mixing. Figure 2d, however, indicates that any possible phonon-induced spin dephasing time will have to be much longer than $T_2$* induced by the nuclear spin bath. It should be noted that earlier spin dephasing and lifetime studies at 100 mK used a magnetic field along the SiV axis to minimize effects of spin-state mixing [19]. In addition, spin dephasing and spin lifetime in SiV centers have also been investigated in HPHT bulk diamond containing a large concentration of substitutional nitrogen impurities. This study showed a $T_2$* about 29 ns at 40 mK and $T_1^{spin}$ about 9 µs at 0.8 K [15], which differ significantly from the properties of SiV centers in high purity CVD-grown diamond reported in this work as well as earlier experimental studies[13, 14, 16].

To probe phonon-induced spin relaxation, we have measured the spin lifetime, $T_1^{spin}$, using the initialization-readout approach developed in an earlier study[13]. As illustrated in Fig. 3a, a laser pulse resonant with the spin-conserved $C_2$ transition excites the SiV center and initializes it into the spin-up state, $|e_+\uparrow>'_g$, through an optical pumping process, specifically, decay via the spin-flip $C_4$ transition. A second laser pulse, again resonant with the $C_2$ transition, arriving after a waiting period, $\tau$, reads out the population in the spin-down state, $|e_-\downarrow\rangle'_g$. This population is in large part due to spin relaxation occurring during the waiting period. Figure 3b shows an example of the time resolved fluorescence obtained in an initialization-readout experiment. The initial decay of the fluorescence during the leading edge of the optical pulses reflects the rapid optical pumping process. Figure 3c plots the ratio of the initial peaks of time resolved florescence observed (the second over the first) as a function of the waiting period. The spin lifetime can be extracted from the exponential recovery of the peak ratio with increasing waiting period.

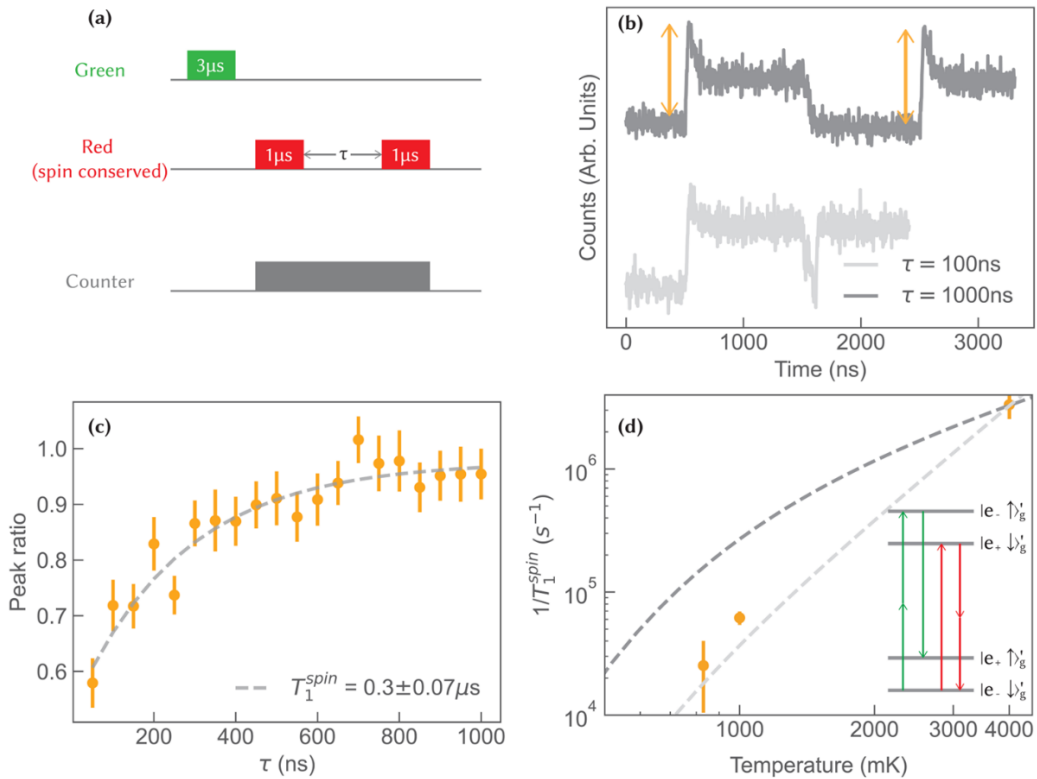

**Fig. 3.** (a) Schematic of the pulse sequence used for the spin lifetime measurement. (b) Fluorescence as a function of time showing the initialization and readout of the SiV spin state at 4 K. Orange arrows denote peak values used for numerical fitting. (c) The peak ratio (the second over the first) as a function of the waiting period, $\tau$, at 4 K. The dashed line is an exponential fit to a recovery time of 0.3 μs. (d) Temperature dependence of the spin lifetime. Dark and light dashed lines are numerical fits to temperature dependence of spin relaxation rates that involve single-phonon spin-flip and two-phonon spin-flip transitions, respectively, as discussed in the text.

The inset shows schematically two relaxation pathways that involve two-phonon spin-flip transitions.

Figure 3d shows $1/T_1^{spin}$ extracted from the initialization-readout experiments at temperatures between 4 K and 830 mK. The spin lifetime obtained near 4 K is in good agreement with earlier studies. At T<500 mK, our experiments indicate $T_1^{spin} > 0.1$ ms (not shown in Fig. 3d). Note that for the spin lifetime measurement, a time resolution capable of resolving the initial optical pumping peak, which occurs in a 10 ns timescale, is needed. This necessitates an extended averaging time. However, as the waiting period increases, the required averaging time becomes increasingly challenging to achieve. In particular, during long averaging periods, the SiV resonance can shift due to spectral diffusion. The charge state may also change due to continuous resonant excitations. These experimental difficulties limited the longest lifetime that can be adequately measured in our current experimental setup to about 0.1 ms.

The nearly 100-fold increase of spin lifetime when the temperature is decreased from 4 K to slightly below 1 K shown in Fig. 3d corresponds to strong suppression of phonon-induced spin relaxation. The steep increase of $T_1^{spin}$ also indicates that the spin relaxation results from transitions between the two lower and upper states in the ground state doublet separated by $\lambda_{so}^g$. Specifically, the spin relaxation can take place in a dynamical process, for which a SiV in a lower state is excited to an upper state through phonon absorption and then returned to the other lower state through phonon emission. Note that the spin-flip can occur via either phonon absorption or phonon emission. In the limit that the transitions between the lower and upper states in the ground state doublet take place via a single-phonon process, i.e., the first order electron-phonon interactions, the overall spin relaxation rate is expected to scale with $n(T)[1+n(T)]$. However, as shown by the numerical fit in Fig. 3d, this temperature dependence can only lead to a 25-fold increase in the spin lifetime when the temperature is decreased from 4 K to 1 K.

The ground-state energy level structure of the SiV center shown in Fig. 1a dictates that spin-flip transitions between the lower and upper states in the ground state doublet take place between two states that have the opposite spin orientations but nearly the same orbital components. As a result, the single-phonon spin-flip transition rate depends not only on the spin-state mixing induced by the off-axis magnetic field, but also on the orbital-state mixing induced by strain and JT effects, because the selection rule for acoustic transitions requires that the single-phonon transition changes the orbital states involved in the transition. In comparison, a two-phonon transition between the lower and upper states does not change the orbital states involved. Because of the relatively small orbital state mixing, the two-phonon spin-flip transition rate can exceed or at least be comparable to the corresponding one-phonon spin-flip transition rate.

The large discrepancy between the observed temperature dependence and the $n(T)[1+n(T)]$ dependence shown in Fig. 3d indicates significant contributions from the two-phonon spin-flip transitions. As shown in the inset of Fig. 3d, the two-phonon spin-flip transition can take place via either phonon absorption or phonon emission. The corresponding spin relaxation rate is expected to scale with $n_{1/2}^2(T)[1+n(T)]$ and $n(T)[1+n_{1/2}(T)]^2$, respectively, where $n_{1/2}(T)$ is the thermal population for phonons with frequency= $\lambda_{so}^g/2$ (it can be shown that $n_{1/2}^2(T)[1+n(T)] = n(T)[1+n_{1/2}(T)]^2$ ). This temperature dependence can provide a good description of the observed temperature dependence, as shown by the numerical fit in Fig. 3d. Nevertheless, a detailed theoretical analysis is still needed to determine the relative contributions of the single-phonon and two-phonon spin-flip transitions, which cannot be reliably extracted from the numerical fit.

In addition to the multi-phonon spin relaxation processes discussed above, spin relaxation can also take place through the direct acoustic transition between the two lower energy ground states with a frequency separation near 3 GHz. The spontaneous emission lifetime for the direct acoustic transition is an important parameter for phononic cavity QED studies. At 1 K, the thermal occupation for phonons near 3 GHz is approximately 6. The observed $T_1^{spin}$ near 1 K thus sets a lower limit of about 0.2 ms for the spontaneous emission lifetime for the direct acoustic transition. The actual spontaneous emission lifetime is expected to be much longer since phonon-induced spin-flip transitions between the lower and upper states of the ground state doublet are expected to play a dominant role near 1 K.

IV. Summary

In summary, our experimental studies of CPT and spin relaxation of a SiV center in a temperature range between 4 K and 100 mK have provided valuable information on both phonon-induced spin dephasing and phonon-induced spin lifetime. The phonon-induced spin phasing arises primarily from orbital relaxation induced by the first order electron-phonon interactions, which is expected from earlier studies at higher temperatures. The temperature dependence of the spin lifetime, however, indicates that second-order electron-phonon interactions, specifically, two-phonon spin-flip transitions, play an essential role in the spin relaxation of the SiV ground states. While there has been no experimental evidence on possible effects of the direct acoustic transition between the two lower energy spin states in spin relaxation, the spin lifetime observed sets a lower limit on the spontaneous emission lifetime for the direct acoustic transition. We hope that these results can prompt a more systematic theoretical analysis of the spin relaxation processes in SiV centers.

**Acknowledgement**

The research effort at the University of Oregon has been supported by NSF under Grant Nos. 2012524 and 2003074 and by the ARO MURI through Grant No. W911NF-18-1-0218. The effort at the Oak Ridge National Laboratory has been supported by the U. S. Department of Energy, Office of Science, Basic Energy Sciences, Materials Sciences and Engineering Division.

**Appendix: Power Broadening of Coherent Population Trapping Spectral Response**

We consider a $\Lambda$-type three-level system, where two nearly resonant optical fields, with frequency $\omega_+$ and $\omega_-$, couple the upper state, $|e\rangle$, to the two lower spin states, $|+\rangle$ and $|-\rangle$, respectively. The Rabi frequency of the two optical fields are $\Omega_+$ and $\Omega_-$, respectively. For our experiments, $\Omega_+$ is compatible to $\Omega_-$. In comparison, textbook treatment of CPT assumes $\Omega_+ \ll \Omega_-$, with the electron initially in state $|+\rangle$.

We choose a rotating frame, in which the wave function of the three-level system can be written as

$$|\Psi\rangle = \tilde{C}_+ \exp(i\omega_+ t)|+\rangle + \tilde{C}_e |e\rangle + \tilde{C}_- \exp(i\omega_- t)|-\rangle$$

The density matrix elements in the rotation frame are thus defined as $\rho_{ij} = \langle \tilde{C}_i \tilde{C}_j^* \rangle$. The corresponding density matrix equations with phenomenological decay rates are

$$\dot{\rho}_{e+} = -(i\Delta_+ + \gamma)\rho_{e+} + \frac{i\Omega_+}{2}(\rho_{ee} - \rho_{++}) - \frac{i\Omega_-}{2}\rho_{-+} \qquad (a)$$

$$\dot{\rho}_{e-} = -(i\Delta_- + \gamma)\rho_{e-} + \frac{i\Omega_-}{2}(\rho_{ee} - \rho_{--}) - \frac{i\Omega_+}{2}\rho_{+-} \qquad (b)$$

$$\dot{\rho}_{-+} = -[i(\omega_B - \delta) + \gamma_s]\rho_{-+} + \frac{i\Omega_+}{2}\rho_{-e} - \frac{i\Omega_-}{2}\rho_{e+} \qquad (c)$$

$$\dot{\rho}_{ee} = -\Gamma\rho_{ee} + (\frac{i\Omega_+}{2}\rho_{e+} + c.c.) + (\frac{i\Omega_-}{2}\rho_{e-} + c.c.) \qquad (d)$$

where $\gamma_s$ and $\gamma$ are the decay rates for the spin coherence and the optical dipole coherence, respectively, $\Gamma$ is the decay rate for the excited state population, $\Delta_+ = \omega_0 - \omega_+$, $\Delta_- = \omega_0 - \omega_B - \omega_-$, $\delta = \omega_+ - \omega_-$, with $\omega_0$ and $\omega_B$ being the frequency separation between states $|e\rangle$ and $|+\rangle$ and between states $|+\rangle$ and $|-\rangle$, respectively.

In the limit that $\gamma_s \ll (\gamma, \Gamma)$, which is satisfied in most CPT experiments, the excited-state population and the optical dipole coherence described by $\rho_{e+}$ and $\rho_{e-}$ can reach steady state in a timescale much faster than that for the spin coherence described by $\rho_{-+}$. In this case, $\rho_{e+}$ and $\rho_{e-}$ as well as the diagonal matrix elements follow adiabatically the dynamics of $\rho_{-+}$, with

$$\rho_{e+} = -\frac{i}{2\gamma}(\Omega_+ N_+ + \Omega_- \rho_{-+})$$

$$\rho_{e-} = -\frac{i}{2\gamma}(\Omega_- N_- + \Omega_+ \rho_{+-})$$

Where $N_\pm = \rho_{\pm\pm} - \rho_{ee}$ is the population difference between the corresponding lower and upper states. In addition, we have also assumed $|\Delta_\pm| \ll \gamma$ and thus have set $\Delta_\pm = 0$ in Eq. **. The steady-state excited-state population is then given by

$$\rho_{ee} = \frac{1}{2\Gamma\gamma}[(\Omega_+^2 N_+ + \Omega_-^2 N_-) + 2\Omega_+\Omega_- \operatorname{Re}(\rho_{-+})].$$

Using the above results and Eq. **, we then arrive at the equation of motion for the spin coherence

$$\dot{\rho}_{-+} = -[i(\omega_B - \delta) + \gamma_s + \frac{\Omega_+^2 + \Omega_-^2}{4\gamma}]\rho_{-+} - \frac{\Omega_+\Omega_-}{4\gamma}(N_+ + N_-).$$

The steady-state solution of $\rho_{-+}$ is thus given by

$$\rho_{-+} = -\frac{\Omega_+\Omega_-}{4\gamma}\frac{N_+ + N_-}{i(\omega_B - \delta) + \gamma_s + (\Omega_+^2 + \Omega_-^2)/4\gamma}.$$

The $\Omega_\pm^2/4\gamma$ terms in Eq. (*) correspond to the power broadening of the optically driven spin transition and thus the power broadening of the CPT resonance. As shown by Eq. *, the CPT spectral response is determined by the real part of $\rho_{-+}$.